# Structural engineering of transition-metal nitrides for surface-enhanced Raman scattering chips

Leilei Lan, Haorun Yao, Guoqun Li, Xingce Fan, Mingze Li, and Teng Qiu (✉)

School of Physics, Southeast University, Nanjing 211189, China

https://doi.org/10.1007/s12274-021-3904-z

https://www.researchgate.net/profile/Leilei-Lan-2

# Structural engineering of transition-metal nitrides for surface-enhanced Raman scattering chips

Leilei Lan[1], Haorun Yao[1], Guoqun Li[1], Xingce Fan[1], Mingze Li[1], and Teng Qiu[1]*

[1] School of Physics, Southeast University, Nanjing 211189, China

E-mail: tqiu@seu.edu.cn

## Abstract

Noble-metal-free surface-enhanced Raman scattering (SERS) substrates have attracted great attention for their abundant sources, good signal uniformity, superior biocompatibility, and high chemical stability. However, the lack of controllable synthesis and fabrication of noble-metal-free substrates with high SERS activity impedes their practical applications. Herein, we propose a general strategy to fabricate a series of planar transition-metal nitride (TMN) SERS chips *via* an ambient temperature sputtering deposition route. These planar TMN (tungsten nitride, tantalum nitride, and molybdenum nitride) chips show remarkable Raman enhancement factors (EFs) with ~$10^5$ owing to efficient photoinduced charge transfer process between TMN chips and probe molecules. Further, structural engineering of these TMN chips is used to improve their SERS activity. Benefiting from the synergistic effect of charge transfer process and electric field enhancement by constructing nanocavity structure, the Raman EF of WN nanocavity chips could be greatly improved to ~$1.29 \times 10^7$, which is an order of magnitude higher than that of planar chips. Moreover, we also design the WN/monolayer MoS$_2$ heterostructure chips. With the increase of surface electron density on the upper WN and more exciton resonance transitions in the heterostructure, a ~$1.94 \times 10^7$ level EF and a $5 \times 10^{-10}$ M level detention limit could be achieved. Our results provide important guidance for the structural design of ultrasensitive noble-metal-free SERS chips.

# 1 Introduction

As a powerful analytical tool, surface-enhanced Raman scattering (SERS) has been widely applied in the fields of biomedicine, food safety, environmental monitoring, and reaction process analysis for its high sensitivity, nondestructive detection, and spectroscopic fingerprinting [1–5]. So far, as the most conventional material for SERS substrates, noble metals show remarkable SERS activity because of their localized surface plasmon resonance (LSPR) effect based on electromagnetic mechanism (EM) [6]. Nonetheless, noble-metal SERS substrates carry several inherent shortcomings, restricting their further applications. For example, the materials used for noble-metal SERS substrates are mostly limited to Au and Ag, which confines the material resources. Besides, for Au and Ag, the former is costly, while the latter lacks biological compatibility, and is likely to lose SERS activity once vulcanized by sulfide from environment or oxidized by Raman laser irradiation [7]. Therefore, it is of great significance to develop novel SERS materials to replace traditional noble metals.

In recent years, searching for noble-metal-free SERS materials with high sensitivity,

superior stability, and low-cost has attracted much attention. The SERS activity of noble-metal-free materials mainly originates from the photoinduced charge transfer (PICT) process between the SERS materials and probe molecules, that is the chemical mechanism (CM) [8, 9]. By optimizing the morphology [10, 11], crystallinity [12–14], crystal orientation [15, 16], stoichiometric ratio [17, 18], thickness [19-21], pressure [22, 23] of noble-metal-free materials, the polarizability of probe molecules would be significantly enhanced, which could give these materials noble-metal-comparable SERS enhancement. Until now, a series of novel noble-metal-free materials such as transition-metal dichalcogenides [24, 25], transition-metal oxides [26, 27], two-dimensional (2D) materials [28, 29], organic semiconductors [30, 31], metal organic frameworks (MOFs) [32], and perovskites [33] have been continually researched and reported. However, when these materials are used to fabricate SERS substrates, some problems are exposed. For instance, transition-metal oxides require to make oxide vacancies by complicated defect engineering. 2D materials and organic semiconductors are expensive. The detection ability of perovskites is weak. And it is difficult to planarize the surface of MOFs nanopowder substrates, resulting in their poor signal uniformity. In addition, obtaining noble-metal-free materials with the LSPR effect is another strategy to displace noble metals and fabricate sensitive SERS substrates. However, the plasmon resonance peaks of most noble-metal-free materials are located in the infrared region because of their low free carrier concentrations. Fortunately, some methods such as careful doping, adjustment of morphology or size have recently been used to obtain the LSPR effect in the visible range [34–36]. For instance, the plasmonic $WO_{3-x}$ [37, 38], $MoO_{3-x}$ [39], $WC_{0.82}$ [40], and $Ti_3C_2$ [41] possess excellent SERS performances under the irradiation of visible laser. However, being similar to the noble metals, these plasmonic noble-metal-free materials also need accurate regulation in the gap, shape, and size of structural units to improve the 'hot spots' density and achieve high SERS activity. Besides, the metal oxides require precise control of oxygen vacancies to realize the LSPR effect in the visible range. Furthermore, most noble-metal-free SERS materials are usually prepared by complex chemical synthesis methods, making it difficult to achieve planarization and synthesize various materials. Thus, it is still a great challenge to develop a simple, low-cost, and universal method to fabricate the noble-metal-free SERS materials with excellent performances.

With the advantages of excellent conductivity, high hardness, and superior stability, transition-metal nitrides (TMNs) are widely applied in catalysis, energy storage, and medicine [42–45]. In recent years, TMNs have also been reported to be the SERS substrates [46–50]. For instance, Guan *et al.* reported that ultrathin δ-MoN nanosheets were synthesized *via* a solution route at 270 °C and 12 atm, and the Raman EF of these nanosheets is determined to $8.16 \times 10^6$ [46]. George *et al.* demonstrated that few-layered $Ti_2N$ MXenes exhibited ultrasensitive detection of explosives at micromolar level [47]. Xi *et al.* developed a molten-salt method for preparing plasmonic vanadium nitride porous materials, which could be used as sensitive and stable SERS substrates [48]. Nevertheless, to the best of our knowledge, the current study is almost limited to titanium nitride, molybdenum nitride and vanadium nitride, the systematic preparation and research of other TMNs have not been reported, besides the Raman enhancement mechanism is still unclear. In addition, because most of TMNs are inert in thermodynamics under an ambient environment, thereby these reported TMN SERS materials usually require harsh chemical syntheses condition such as high temperature and high pressure, which limits their security and low-cost preparation. Furthermore, these reported TMN SERS materials are always prepared as nanopowders rather than chips, resulting in complex pretreatment before SERS detection.

Herein, a simple and general method to fabricate a series of planar TMN SERS chips are demonstrated, in which the WN and TaN are taken as SERS materials for the first time. By means of experimental and theoretical research, we prove that sensitive SERS activity of these planar TMN chips is attributed to the efficient PICT process between TMN chips and probe molecules, and their Raman enhancement factors (EFs) are up to $\sim 10^5$. Furthermore, by constructing nanocavity structure, the Raman EF could be increased to ~1.29 × $10^7$, which is an order of magnitude higher that of planar chips. We also design the WN/monolayer $MoS_2$ heterostructure chips, achieving a ~1.94 × $10^7$ level EF and 5 × $10^{-10}$ M level detention limit. Our research results provide economical and effective guidance for the structural design of ultrasensitive noble-metal-free SERS chips.

## 2 Results and Discussion

### 2.1 Characterization of TMN chips

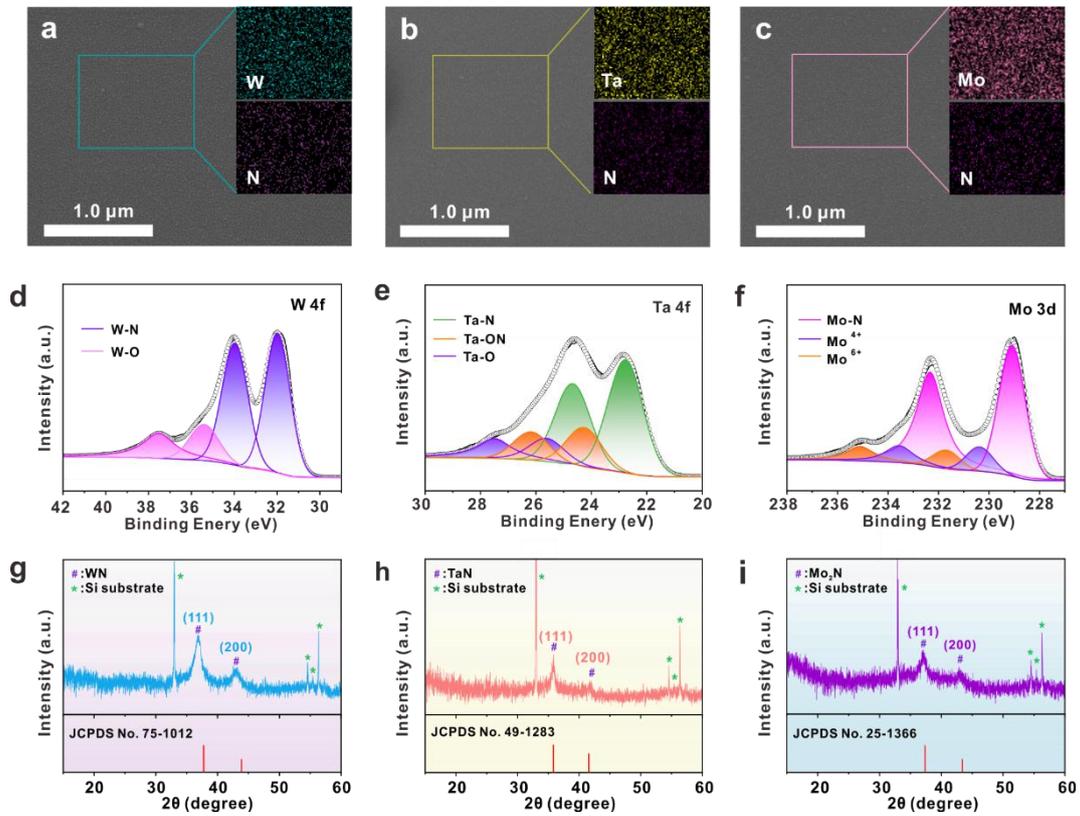

**Figure 1** (a)–(c) The SEM images and elemental mapping images of (a) WN, (b) TaN and (c) $Mo_2N$ chips. (d)–(f) The W 4f, Ta 4f and Mo 3d XPS spectra of (d) WN, (e) TaN and (f) $Mo_2N$ chips. (g)–(i) The XRD patterns of (g) WN, (h) TaN and (i) $Mo_2N$ chips.

Tungsten nitride, tantalum nitride, and molybdenum nitride are deposited on Si wafer using direction current (DC) reactive magnetron sputtering technique, widely used for the large-scale production of chips. The morphologies of these sputtered TMN chips are investigated by scanning electron microscopy (SEM). As shown in Figs. 1(a), 1(b) and 1(c), these TMN chips present uniform and density surface morphologies. In the elemental mapping analysis (insets of the Figs. 1(a), 1(b) and 1(c)), the W, Ta, Mo and N elements that are uniformly dispersed also testifies the structural uniform of these TMN chips, which is essential for homogeneous distribution of SERS-active points in these chips. Further, the chemical states of these TMN chips are analyzed by X-ray photoelectron spectroscopy (XPS). As shown in Fig. 1(d), typical three-peak-shaped W4f XPS spectrum can be well fitted into two spin-orbit doublets, among which the doublet peaks located at 31.9 and 34.0 eV are ascribed to W-N chemical bond, while the two small peaks at the high binding energy range (35.4 and 37.5 eV) are corresponding to W-O chemical bond [51]. It should be pointed out that the oxygen signal should result from the unavoidable surface oxidation [52]. In the Ta 4f XPS spectrum of tantalum nitride (Fig. 1(e)), the spectrum can be deconvoluted into three doublets. The two characteristic strong peaks located at 22.8 and 24.7 eV are associated with Ta-N bond. The peaks located at 24.2 and 26.1 eV are assigned to the tantalum oxynitride (Ta-ON), while the peaks located at 25.6 and 27.5 eV match well with Ta-O bond [53]. As shown in the Fig. 1(f), the Mo3d spectrum can be fitted into three spin-orbit doublets. The main doublet peaks at the binding energy of 229.1 and 232.3 eV are defined as the Mo-N bond in molybdenum nitride. The small peaks located at the binding energy of 230.5, 233.8, 231.8 and 235.2 eV are classified as $Mo^{4+}$ $3d_{5/2}$, $Mo^{4+}$ $3d_{3/2}$, $Mo^{6+}$ $3d_{5/2}$ and $Mo^{6+}$ $3d_{3/2}$, respectively [54]. All the XPS measurements demonstrate the presence of the dominant TMN components in these sputtered chips. In addition, X-ray diffraction (XRD) patterns are collected to evaluate the crystal structure of these TMN chips. In the XRD pattern of tungsten nitride (Fig. 1(g)), two diffraction peaks at 36.8° and 42.9° could be clearly observed, which are assigned to the (111) and (200) planes of the cubic WN (JCPDS No. 75-1012), respectively. The rest of diffraction peaks labeled as star (*) are from the Si substrate (Fig. S1 in the Electronic Supplementary Material (ESM)). In the Figs. 1(h) and 1(i), the XRD patterns of tantalum nitride and molybdenum nitride chips could be accurately indexed as cubic TaN (JCPDS No. 49-1283) and cubic $Mo_2N$ (JCPDS No.

25-1366), respectively. Thus, we could confirm that these planar TMN chips have been prepared successfully.

## 2.2 SERS properties of TMN chips

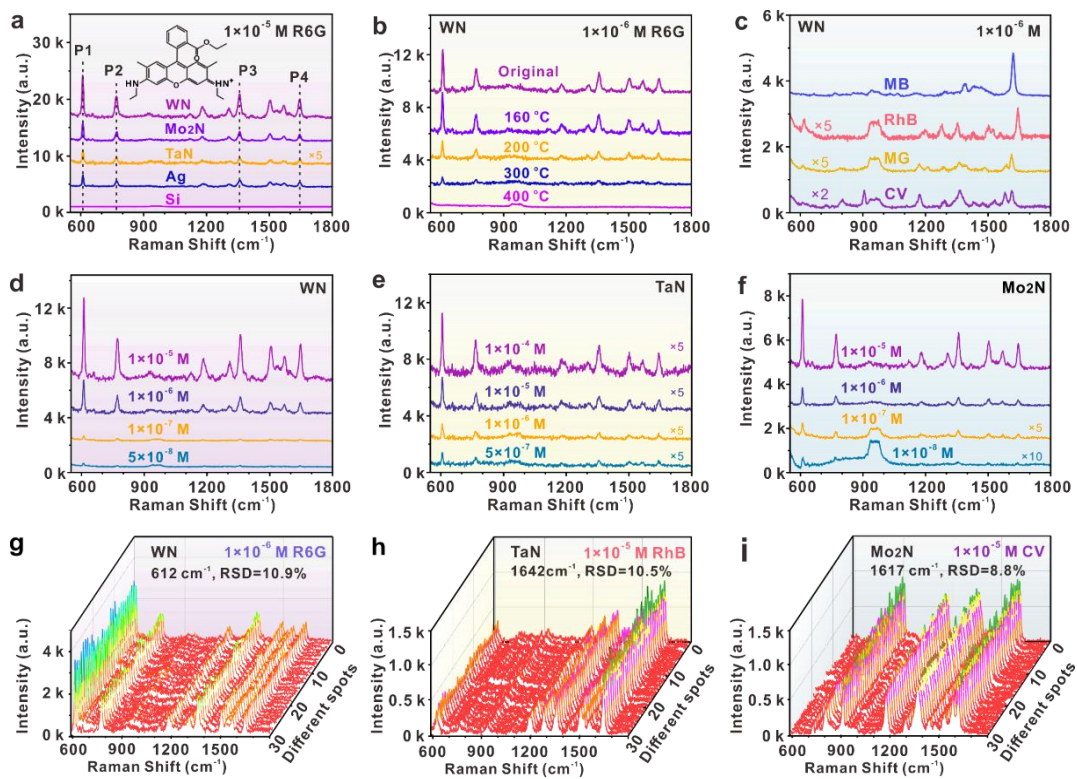

**Figure 2** (a) The Raman spectra of R6G ($10^{-5}$ M) adsorbed on bare Si, Ag, WN, TaN and Mo$_2$N chips. (b) The Raman spectra of R6G ($10^{-6}$ M) gathered from WN chips under high temperature treatment. (c) The Raman spectra of a series of organic compounds ($10^{-6}$ M) adsorbed on WN chips. (d)–(f) The Raman spectra of R6G with different concentrations adsorbed on (d) WN, (e) TaN and (f) Mo$_2$N chips. (g)–(i) The Raman spectra of R6G ($10^{-6}$ M), RhB ($10^{-5}$ M) and CV ($10^{-5}$ M) gathered from 30 random sites on the (g) WN, (h) TaN and (i) Mo$_2$N chips.

To investigate the SERS performances of the WN, TaN and Mo$_2$N chips, a typical molecule R6G is chosen as the Raman probe. Fig. 2(a) shows the Raman spectra of R6G ($10^{-5}$ M) adsorbed on the bare Si substrate, Ag (the preparation and characterizations of Ag chips are shown in Section S1 and Fig. S2 in the ESM), WN, TaN and Mo$_2$N chips. Obviously, no detectable Raman signals related with R6G molecule could be found on bare Si substrate, excluding the effect of the Si substrate in the SERS action. However, four main Raman characteristic peaks of R6G at P1 (612 cm$^{-1}$), P2 (773 cm$^{-1}$), P3 (1360 cm$^{-1}$) and P4 (1650 cm$^{-1}$) are clearly displayed on the Raman spectra acquired on these TMN chips. The peaks located at P1 and P2 can be referred to the bending motions of C and H atoms of the xanthene skeleton, respectively. The other two peaks (P3 and P4) are assigned the C−C stretching vibrations of the aromatic nucleus [27]. To the best of our knowledge, it is the first time for WN and TaN to be reported as the SERS materials. Furthermore, it could note that the Raman signal intensities of R6G molecules absorbed on WN and Mo$_2$N chips are even higher than those on Ag chips, displaying the remarkable SERS activity of these TMN chips. In addition, these TMN SERS chips also possess high stability, even if they are heated 160 $^\circ$C in air for 30 min, their sensitivity almost unchanged (Fig. 2(b) and Fig. S3 in the ESM). To demonstrate the universality of these TMN chips, we also test other dyes such as crystal violet (CV), malachite green (MG), rhodamine B (RhB), and methylene blue (MB). These dyes could be effectively detected on these TMN SERS chips at the level below than 10$^{-5}$ M (Fig. 2(c) and Fig. S4 in the ESM), which indicates the huge potential of these chips for chemical sensing. Further, SERS measurements are performed under different concentrations of R6G. As shown in Figs. 2(d), 2(e) and 2(f), WN, TaN and Mo$_2$N chips could detect R6G at the concentrations of 5 × 10$^{-8}$ M, 5 × 10$^{-7}$ M and 1 × 10$^{-8}$ M, respectively. Based on the standard equation (detailed calculations of the EF are presented in Section S2 in the ESM), the EFs of WN, TaN and Mo$_2$N chips are calculated to be ~6.77 × 10$^5$ at the R6G concentration of 5 × 10$^{-8}$ M, ~4.62 × 10$^4$ at the R6G concentration of 5 × 10$^{-7}$ M and ~6.50 × 10$^5$ at the R6G concentration of 1 × 10$^{-8}$ M, respectively. In order to further show the signal uniformity of these TMN chips, we detect the Raman signals of 10$^{-6}$ M R6G adsorbed on WN, 10$^{-5}$ M RhB adsorbed on TaN and 10$^{-5}$ M CV adsorbed on Mo$_2$N chips, respectively. As shown in Figs. 2(g), 2(h), 2(i) and S5 in the ESM, the Raman spectra obtained from 30 random points are applied to calculate their relative standard deviation (RSD). The RSD values calculated for the characteristic peaks at 612 cm$^{-1}$ (10$^{-6}$ M R6G adsorbed on WN), 1642 cm$^{-1}$ (10$^{-5}$ M RhB adsorbed on TaN) and 1617 cm$^{-1}$ (10$^{-5}$ M CV adsorbed on Mo$_2$N) are 10.9%, 10.5% and 8.8%, respectively, which proves the excellent signal uniformity of these TMN chips. In a word, all these results clearly confirm that these TMN chips possess outstanding performances as SERS materials.

**2.3 Enhancement mechanism of TMN SERS chips**

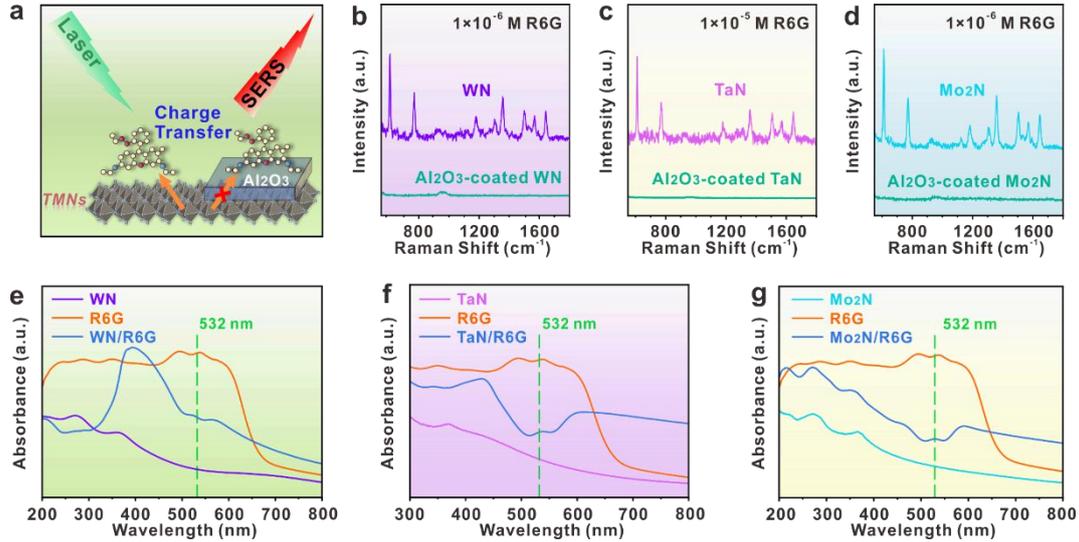

**Figure 3** (a) Schematic illustration of the experiment settings for analyzing the Raman enhancement mechanism of probe molecules absorbed on TMN chips. (b)–(d) The Raman spectra of R6G adsorbed on TMNs and Al$_2$O$_3$-coated TMNs chips: (b) WN, (c) TaN (d) Mo$_2$N. (e)–(g) The UV-vis absorption spectra of TMNs, R6G and TMNs-R6G chips: (e) WN, (f) TaN and (g) Mo$_2$N.

In order to clarify the Raman enhancement mechanism of these TMN chips, a dielectric layer of 5 nm Al$_2$O$_3$ is coated on TMN chips by atomic layer deposition (ALD) technique (Fig. 3(a)), which can block the charge transfer (CT) between the SERS substrates and adsorbed molecules [20]. It is interesting that the Raman signals of probe molecules are vanished on Al$_2$O$_3$-coated TMN chips (Figs. 3(b), 3(c) and 3(d)), which demonstrates the CM is the enhancement mechanism. Moreover, the UV-vis absorption spectra of TMNs, R6G and TMNs-R6G are shown in Figs. 3(e), 3(f) and 3(g). Obviously, no absorption peaks in visible range could be observed in the UV-vis absorption spectra of these TMN chips, which also demonstrates that there is not electromagnetic enhancement to these TMN chips. Furthermore, compared with the UV-vis absorption spectrum of R6G, new absorption peaks of TMNs-R6G are dramatically increased in intensity. The obvious changes in absorption spectra reveal that there is a strong interfacial interaction between TMNs and R6G, which indicates interfacial CT processes exist in TMNs-R6G systems [55]. The degrees of CT for R6G ($1\times10^{-5}$ M) on WN, TaN and $Mo_2N$ are 0.65, 0.71 and 0.66, respectively, which further demonstrates that the CM is the main enhancement mechanism (detailed calculations are presented in Section S3 in the ESM). Moreover, the band structure, density of states (DOS), and work function of these TMNs are further calculated *via* the first-principles calculations (detailed calculations are presented in Section S4 in the ESM). As shown in Figs. 4(a), 4(b) and 4(c), the calculated band structures demonstrate that these TMNs show the metallic behaviors with no obvious gap between the valence band (VB) and the conduction band (CB). In addition, the DOS values of WN, TaN, and $Mo_2N$ at Fermi level are 5.24, 0.17 and 4.44 states $eV^{-1}$, respectively (Figs. 4(d), 4(e) and 4(f)), which also reveal the metallic nature of these TMNs. The metallic contribution of WN, TaN, and $Mo_2N$ may result from the partially occupied W 5d, Ta 5d, and Mo 4d states, respectively. The electron transition probability rate ($w_{ab}$) can be expressed by the Fermi's golden rule: $w_{ab} = \frac{2\pi}{\hbar}|\langle b|M|a\rangle|^2 g_b$, where $|a\rangle$ and $|b\rangle$ are the initial and final states for the charge transition, $M$ is the interaction operator of the two states, $\hbar$ is the reduced Planck constant, and $g_b$ is the DOS of the final state [56]. According to the Fermi's golden rule, electron transition probability is linearly correlated with the DOS near the Fermi level during the CT process. Hence, the large numbers of allowed energy states near the Fermi level for PICT in these TMNs can give rise to the high charge transition probabilities, leading to a remarkable Raman enhancement effect. Owing to more abundant electronic states, WN endows better Raman enhancement than TaN and $Mo_2N$, which is consistent with our experimental SERS performances (Fig. 2(a)). Furthermore, the PICT pathways in TMNs-R6G systems are investigated. The highest occupied molecule orbital (HOMO) and the lowest unoccupied molecule orbital (LUMO) of R6G probe are -5.70 eV and -3.40 eV, respectively. According to the DFT calculated results of work functions, the Fermi level of WN, TaN, and $Mo_2N$ are located at -4.88, -5.27 and -3.72 eV, respectively (Fig. S6, ESM). In general, the polarization tensor (α) can be expressed as α = A + B + C, where A stems from the resonance Raman scattering, B and C are relevant to the substrate-to-molecule and molecule-to-substrate CT transitions [57]. Under irradiation with 532 nm (2.33 eV), the possible PICT pathways between TMNs and R6G are plotted in Figs. 4(g), 4(h), and 4(i). In TMNs-R6G systems, both the PICT processes from the R6G HOMO to the TMNs Fermi level and the TMNs Fermi level to the R6G LUMO in TMNs-R6G systems are contributed to the Raman enhancement [57]. What's more, the molecular resonance could be achieved because the laser excitation energy is close to the electronic transition energy of R6G [9], further magnifying the polarization tensor of the molecules. In addition, XPS characterizations and poor crystallinities indicate that the defects may exist in these chips (Fig. 1) (58). Many researches have reported that TMN materials still exhibit metallic properties in existence of defects [59-61]. We surmise that proper introduction of defects may improve the DOS near Fermi surface, further improve the SERS activity of TMN materials.

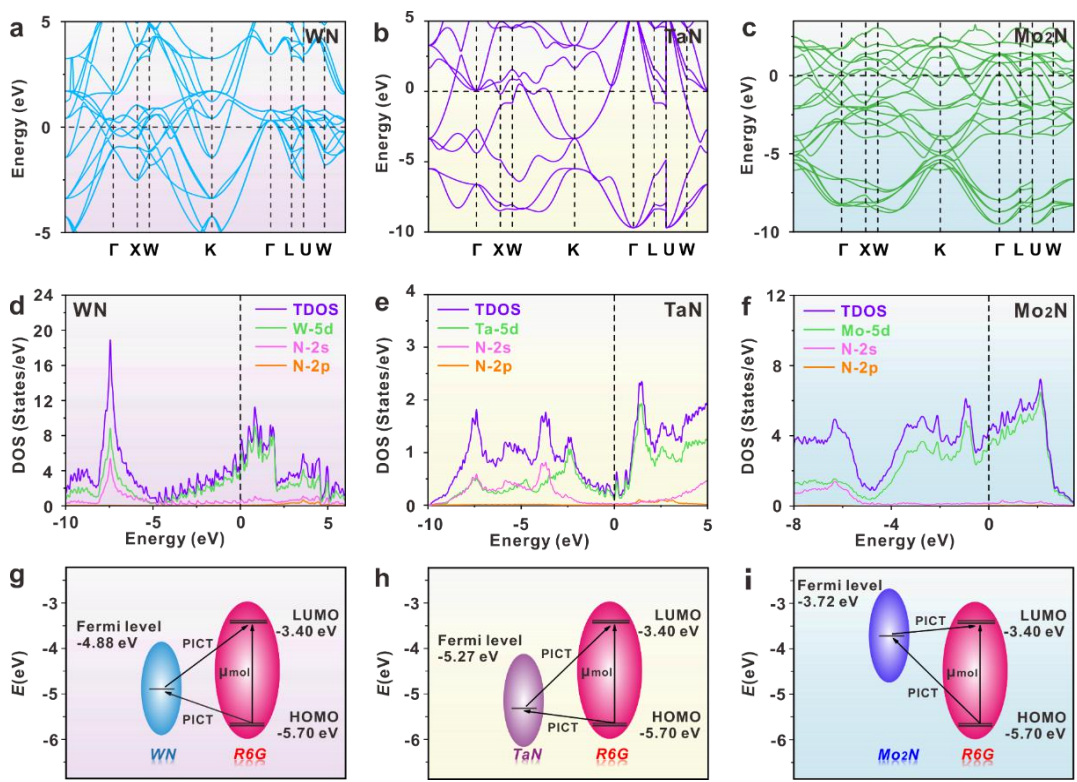

**Figure 4** (a)–(c) The DFT calculated band structure of cubic (a) WN, (b) TaN and (c)

$Mo_2N$. (d)–(f) The DFT calculated DOS of cubic (d) WN, (e) TaN and (f) $Mo_2N$. (g)–(i) The schematic energy level diagrams of (g) WN-R6G, (h) TaN-R6G and (i) $Mo_2N$-R6G systems.

**2.4 SERS properties of TMN-based nanocavity chips**

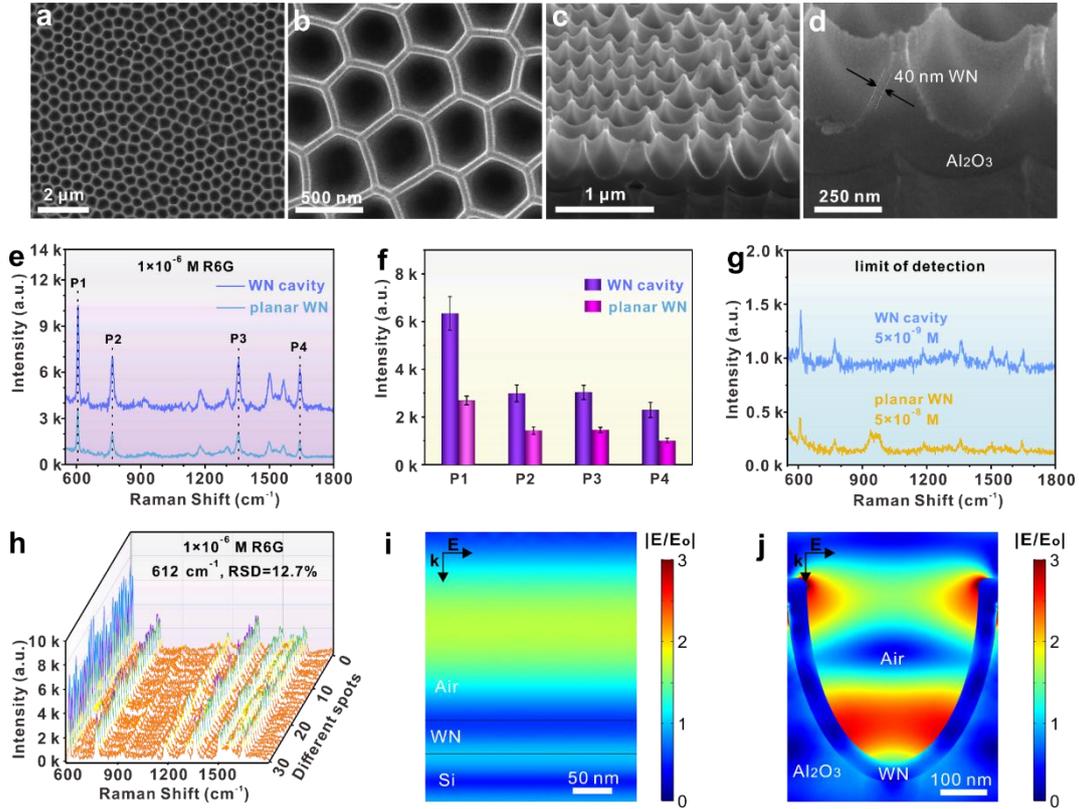

**Figure 5** (a)–(d) The SEM images of WN nanocavity chips: (a) and (b) The top view images, (c) and (d) the cross-sectional images. (e) and (f) The Raman spectra and corresponding signal intensity of $10^{-6}$ M R6G adsorbed on WN nanocavity and planar chips. (g) Comparison of detection limit of WN nanocavity and planar chips. (h) Raman spectra of $10^{-6}$ M R6G gathered from 30 random sites on the WN nanocavity chips. (i) and (j) The simulated electric field distribution in (i) planar and (j) nanocavity-structured WN chips.

In order to further improve the SERS properties of TMN chips, we design the nanocavity arrays based on TMN materials. Taking WN as an example, the WN nanocavity arrays were obtained by sputtering WN thin layer on the V-shaped anodized aluminum oxide (AAO) templates. From the SEM images (Fig. 5(a), 5(b), 5(c) and 5(d)), we could observe that the WN nanocavities orderly arrange into hexagonal arrays with a periodicity of ~400 nm，while the average depth of cavities and the thickness of WN thin layer are ~400 nm and ~40 nm respectively. To evaluate the role of nanocavities in SERS enhancement, Raman spectra of R6G ($10^{-6}$ M) on the WN nanocavity and planar chips are measured and shown in Fig. 5(e). It is obvious that the SERS signals observed from the WN nanocavity chips are much higher than that of planar chips. By monitoring P1, the WN nanocavity chips show ~2.2 times Raman intensity compared with planar chips (Fig. 5(f)). Further, the WN nanocavity chips exhibit an ultrasensitive detection limit of $5 \times 10^{-9}$ M for R6G, while planar chips only show the detection limit of $5 \times 10^{-8}$ M for R6G (Fig. 5(g)). The Raman EF of WN nanocavity chips is calculated to be ~$1.29 \times 10^7$ at R6G concentration of $5 \times 10^{-9}$ M, which is an order of magnitude higher than that of planar chips (~$6.77 \times 10^5$). The above results demonstrate that the SERS effect of WN chips could be further improved by constructing nanocavity array structure. Fig. 5(h) shows the Raman spectra of R6G at 30 random sites on the WN nanocavity chips. The RSD value calculated for the characteristic peaks at 612 $cm^{-1}$ is 12.7%, which demonstrates the good signal uniformity of WN nanocavity chips (Fig. S7 in the ESM). To explain the enhancement phenomenon, simulated electric field distribution of the WN chips is executed by the finite element method simulations (more simulation details can be found in Section S5 and Fig. S8 in the ESM). For the planar WN chips, the electric field distribution is very homogeneous near the WN surface, and the intensity has not enhanced obviously (Fig. 5(i)). Because the incident direction perpendicular to the planar WN surface in this simulation, this uniform electric field distribution is mainly caused by the specular reflection of the planar WN layer. For the WN nanocavity chips, the electric field could be enhanced and the maximum electrical field is located at the sidewall of nanocavities (Fig. 5(j)), indicating that significant SERS enhancement could be obtained for the adsorbed molecules on the cavity walls. The maximum electric field intensities ($|E|/|E_0|$) are 6.32 and 1.78 for the WN nanocavity and planar chips, respectively. The SERS signals are approximately proportional to the fourth power of the enhancement of the local electric field [62]. Thus, the ratio of Raman EF obtained from simulated maximum $|E|/|E_0|$ for the WN nanocavity and planar chips is 159, which is larger than the experimental value ($1.29 \times 10^7/6.77 \times 10^5 = 19$) because the electric field distribution is not homogeneous near the cavity walls and only partial adsorbed molecules are affected by maximum electric field. Benefiting from the synergistic effect of the CT process and electric field enhancement, the final SERS substrates based on nanocavity arrays show an ultralow detection limit. Therefore, structural design of nanocavity is a promising strategy for realizing ultrasensitive SERS sensing.

**2.5 SERS properties of TMN-based heterostructure chips**

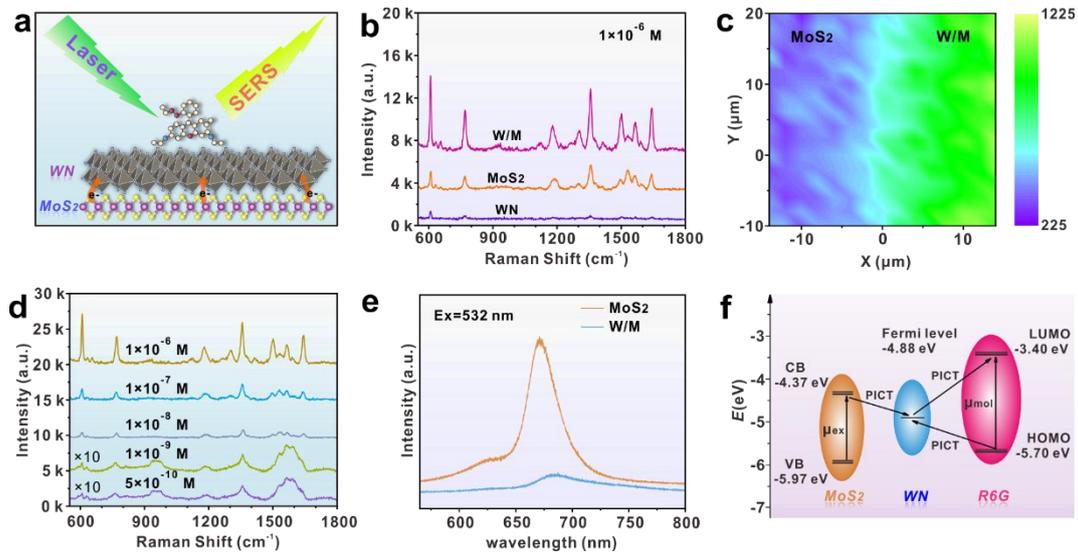

**Figure 6** (a) Schematic illustration of measuring Raman enhancement for probe molecules on the W/M heterostructure chips. (b) Raman spectra of R6G ($10^{-6}$ M) adsorbed on WN, monolayer $MoS_2$, and W/M chips. (c) The Raman mapping image of $10^{-6}$ M R6G (612 cm$^{-1}$) near the dividing line between the monolayer $MoS_2$ part (left) and W/M part (right). (d) Raman spectra of R6G with different concentrations adsorbed on the W/M chips. (e) PL spectra of monolayer $MoS_2$ and W/M chips. (f) The energy level diagram and CT transitions in the R6G-W/M ternary system.

We also design TMN-based heterostructure chips. Taking WN as an example, the WN/monolayer $MoS_2$ (W/M) vertical heterostructures are fabricated by coating a sputtered WN thin layer (~10 nm) on chemical vapor deposition-grown monolayer MoS$_2$. Raman spectra of MoS$_2$ shows that the Raman modes of $E_{2g}^1$ and A$_{1g}$ are located at 385 and 404 cm$^{-1}$ respectively (Fig. S9 in the ESM), which is the characteristic of monolayer MoS$_2$ [63]. Atomic force microscopy (AFM) characterization also proves that the thickness of MoS$_2$ is monolayer-thick with ~0.7 nm (Fig. S10 in the ESM). In the optical photograph of W/M chips, we could clearly see that there is a dividing line between MoS$_2$ and W/M heterostructures (Fig. S11 in the ESM). Further, the elemental mapping analyses demonstrate that the distribution of Mo, S, W and N elements in the W/M heterostructures is very uniform (Fig. S12 in the ESM). All these characterization results confirm that the W/M heterostructure chips have been prepared successfully. Schematic illustration of measuring Raman enhancement for probe molecules on the W/M heterostructures is shown in Fig. 6(a). To examine the effect of W/M chips on the Raman enhancement, the R6G probe molecule is used to investigate the SERS signals of WN, monolayer MoS$_2$ and W/M chips. It is obvious that the SERS signals observed from the W/M chips are much higher than that of WN and MoS$_2$ chips (Fig. 6(b)). Besides, the relative intensity among vibrational bands of R6G is changed when measured on these chips. In the Raman spectra of R6G adsorbed on MoS$_2$, W/M and WN, the intensity ratios of P1 and P3 (IP1/IP3) are 0.71, 1.19 and 2.10, respectively. According to the Herzberg-Teller selection rule, if Raman enhancement is caused by CT, the signal of the asymmetric vibration mode of the probe molecule will be stronger than that of the completely symmetric vibration mode [64]. P1 is the non-totally symmetric vibration mode, while P3 is the totally symmetric vibration mode [13]. In addition, Ren *et al.* have reported that adsorption behavior of R6G molecules on SERS substrate surface would impact the relative peak strength of Raman spectra [65, 66]. And it is found that the intensity of P1 would significantly increase when R6G molecules are vertically adsorbed on the surface of SERS substrate through xanthene ring. In reference to MoS$_2$ material, although its enhancement mechanism is CM, the adsorption direction of most R6G molecules may be tilted rather than perpendicular to the MoS$_2$ surface, resulting in a relatively low value of P1. For WN and W/M which also work by CM, their (IP1/IP3) values are more than 1. However, considering that the molecular adsorption of upper WN is affected by the lower MoS$_2$ of W/M, the value (IP1/IP3) of W/M are between WN and MoS2 (IP1/IP3). Therefore, we believe that the difference of (IP3/IP1) values in Fig. 6b is resulted from the combined effect of CT and molecular adsorption behavior. Furthermore, the Raman mapping integrated intensities of the characteristic peak 612 cm$^{-1}$ is shown in Fig. 6(c). It is also could see that the intensities are stronger on W/M than monolayer MoS$_2$, which could prove that heterostructure able to significantly improve SERS performance. The Raman spectra of R6G with different concentrations absorbed on W/M chips are given in Fig. 6(d). The SERS signals are still conspicuous even when the concentration of R6G solution is decreased to 5 × 10$^{-10}$ M, which is two orders of magnitude lower than that of planar WN chips (5 × 10$^{-8}$ M). The Raman EF of W/M is calculated to be as high as ~1.94 × 10$^7$. The EM contribution is excluded in the MoS$_2$-induced SERS effect, since the MoS$_2$ is not plasmonic material [25, 67]. To elucidate the Raman enhancement mechanism of W/M chips, the photoluminescence (PL) spectra of monolayer MoS$_2$ and W/M heterostructures are measured under excitation with 532 nm (Fig. 6(e)). We could see that the fluorescence of monolayer MoS$_2$ is quenched and the PL intensity is reduced to 9% of pristine intensity after coating WN layer. Further, the transmittance of (10 nm) WN layer is above 75% in the range of 500-800 nm (Fig. S13 in the ESM). These results prove that the main reason for weak fluorescence signal is not the blocking effect of WN layer (9% << 75%). The fluorescence quenching of MoS$_2$ indicates that the recombination process in the pristine MoS$_2$ changed and the excited electrons tend to migrate to WN rather than back to the ground state directly **[68]**. As a result, there are more electrons could reach the Fermi level of WN and electronic density of WN for transferring to probe molecules is increased. The energy level diagram and CT transitions in the R6G-W/M ternary system is shown in Fig. 6(f), the VB and CB of MoS$_2$ are -5.97 eV and -4.37 eV, respectively [69]. Obviously, the required extra energy between the CB of MoS$_2$ and Fermi level of WN is only 0.51 eV, which also indicates that the electrons could transfer from MoS$_2$ to WN easily under laser excitation [70]. According to the Fermi's golden rule, abundant energy states near the Fermi level for PICT in WN could produce high charge transition probabilities, leading to a significant Raman enhancement effect [62]. Apart from that, there are a molecular resonance in R6G and a new exciton resonance transition in MoS$_2$, thus the stronger CT resonances can be expected through vibronic coupling [8]. These synergistic effects could achieve dramatically enhanced PICT processes, which amplifies Raman scattering cross section significantly.

## 3 Conclusions

In summary, we demonstrate a facile and universal strategy to fabricate a series of planar TMN SERS chips by an ambient temperature sputtering deposition technique. For the first time, WN and TaN are used as SERS materials. Because of highly efficient PICT process between TMNs and probe molecules, the Raman EFs of these planar TMN chips could be to ~$10^5$. Going further, taking WN as an example, we design WN nanocavity chips and WN/monolayer MoS$_2$ heterostructure chips to improve their SERS activity. Benefiting from the synergistic effect of the CT process and electric field enhancement, the Raman EF of nanocavity chips could be greatly enhanced to as high as ~$1.29 \times 10^7$, which is an order of magnitude higher than that of planar chips. Thanks to the increase of surface electron density of the upper WN layer and more exciton resonance transitions in the heterostructure, the detection limit of R6G molecule on the WN/monolayer MoS$_2$ heterostructure is as low as $5 \times 10^{-10}$ M with a Raman EF of ~$1.94 \times 10^7$. Our results provide a promising strategy for the structural design of novel chips with noble-metal-comparable SERS activity.

## 4 Methods

Fabrication of planar TMN chips: Planar WN chips were prepared on Si wafer at room temperature in a mixture of Ar (99.999%, 20 sccm) and N$_2$ (99.999%, 6 sccm) using the DC magnetron sputtering system without post-treatments. Target was 50 mm-diameter high purity metal W (99.999%) plate. The power of magnetron was 60 W with a pressure of 1.75 Pa for 240 s. The deposition rate of WN was ~0.168 nm s$^{-1}$ (Section S6 and Fig. S14 in the ESM). The sputtering parameters of other TMN chips (TaN and Mo$_2$N) were listed in Table S1 in the ESM.

Fabrication of TMN-based nanocavity array chips: Taking WN as an example, WN nanocavity array chips were obtained by sputtering 40 nm-thick WN on V-shaped AAO templates (were purchased from Shenzhen Top Membranes Technology Co., Ltd.).

Fabrication of TMN-based heterostructure chips: The WN/monolayer MoS$_2$ heterostructure chips were prepared through a convenient method. After obtaining monolayer MoS$_2$ on SiO$_2$/Si wafer *via* chemical vapor deposition, we deposited a thin layer of WN on the monolayer MoS$_2$ by magnetron sputtering technique.

Coating Al$_2$O$_3$ on TMN-based chips: Al$_2$O$_3$ layers were coated on as-sputtered TMN chips in an ALD reactor (MNT-100, Wuxi MNT Micro and Nanotech Co., Ltd.) at 150 °C. The Al$_2$O$_3$ precursors, *i.e.*, trimethylaluminum (TMA, maintained at 150 °C) and water (maintained at 40 °C), were alternatively pumped into the reaction chamber using N$_2$ (99.999%, 20 sccm) as the carrier and purge gas. One complete reaction cycle is composed by four steps: (1) TMA reactant was pulsed for 15 ms; (2) the chamber was purified by N$_2$ for 25 s; (3) water vapor was pulsed for 8 ms; and (4) N$_2$ gas was used to purge the chamber for 30 s. The thickness of Al$_2$O$_3$ was controlled by setting the number of reaction cycles. Typically, the deposition rate of this process is about 0.1 nm cycle$^{-1}$, and 5 nm Al$_2$O$_3$ was obtained after 50 ALD cycles.

Characterization: XRD patterns were obtained using a Rigaku Smartlab (3) X-ray diffractometer ($\lambda$ = 0.15406 nm). The SEM images and the elemental mapping images were recorded on a FEIXL-30 SEM equipped with an energy dispersive spectrometer. The XPS measurements were performed by a Thermofisher ESCALAB 250Xi apparatus. The optical absorption spectra were measured by a UV-vis Spectrophotometer (Shimadzu, UV-3600) in the diffuse reflectance mode. The thickness of the samples was obtained by a BRUKER Dimension Icon AFM. The Raman measurements under laser excitation at 532 nm were achieved from a Jobin Yvon Lab RAM HR800 Raman spectrometer. The Raman signal of probe molecules could be detected if the signal-to-noise ratio is more than or equal to 3. We prepared a series of probe solutions with different concentrations, including $1\times10^{-4}$, $5\times10^{-5}$, $1\times10^{-5}$, $5\times10^{-6}$, $1\times10^{-6}$, $5\times10^{-7}$, $1\times10^{-7}$, $5\times10^{-8}$, $1\times10^{-8}$, $5\times10^{-9}$, $1\times10^{-9}$, $5\times10^{-10}$, and $1\times10^{-10}$ M. Different concentrations of probe molecules are adsorbed on SERS substrates from high concentration to low concentration, respectively. If SERS substrate could detect Raman signal of probe molecules with A concentration rather than B concentration (one level lower than that of A), we would regard that the limit of detection for probe molecule is below A concentration. All samples were maintained in probe solution for 1h and taken out, followed by a drying treatment under $N_2$. The laser beam was focused to a spot about 2 μm in diameter with a 50× objective lens. The laser power on the sample surface was 36.5 μW and the spectra were acquired for 5 s unless specified. The Raman mapping spectra were collected from an area 28 ×30 μm$^2$ with 2 s on the heterostructure chips and the step length is 2 μm.

## Acknowledgements

This work was supported by the National Natural Science Foundation of China (Grant No. 11874108).

**Electronic Supplementary Material:** Supplementary material (Including I. The preparation and characterizations of the Ag chips; II. Calculation of the enhancement factor; III. Calculation of the degree of charge transfer; IV. First-principles calculations; V. Simulation of the electromagnetic field distribution; VI. Deposition rate of the WN films; VII. The supplementary characterization of samples; VIII. The sputtering parameters of TaN and $Mo_2N$ chips.) is available in the online version of this article at http://dx.doi.org/10.1007/s12274-***-****-* (automatically inserted by the publisher).

# References

[1] Cialla-May, D.; Zheng, X. S.; Weber, K.; Popp, J. Recent progress in surface-enhanced Raman spectroscopy for biological and biomedical applications: from cells to clinics. *Chem. Soc. Rev* **2017**, *46*, 3945–3961.

[2] Li, J. F.; Huang, Y. F.; Ding, Y.; Yang, Z. L.; Li, S. B.; Zhou, X. S.; Fan F. R.; Zhang, W.; Zhou, Z. Y.; Wu, D. Y.; Ren, B.; Wang, Z. L.; Tian, Z. Q. Shell-isolated nanoparticle-enhanced Raman spectroscopy. *Nature* **2010**, *464*, 392–395.

[3] Wang, X.; Huang, S. C.; Hu, S.; Yan, S.; Ren, B. Fundamental understanding and applications of plasmon-enhanced Raman spectroscopy. *Nat. Rev. Phys.* **2020**, *2*, 253–271.

[4] Shen, J. L.; Su, J.; Yan, J.; Zhao, B.; Wang, D. F.; Wang, S. Y.; Li, K.; Liu, M. M.; He, Y.; Mathur, S.; Fan, C. H.; Song, S. P. Bimetallic nano-mushrooms with DNA-mediated interior nanogaps for high-efficiency SERS signal amplification. *Nano Res.* **2015**, *8*, 731–742.

[5] Hao, Q.; Li, M. Z.; Wang, J. W.; Fan, X. C.; Jiang, J.; Wang, X. X.; Zhu, M. S.; Qiu, T.; Ma, L. B.; Chu, P. K.; Schmidt, O. G. Flexible surface-enhanced Raman scattering chip: a universal platform for real-time interfacial molecular analysis with femtomolar sensitivity. *ACS Appl. Mater. Interfaces* **2020**, *12*, 54174–54180.

[6] Langer, J.; de Aberasturi, D. J., Aizpurua, J.; Alvarez-Puebla, R. A.; Auguié, B.; Baumberg, J. J.; Bazan, G. C.; Bell, S. E. J.; Boisen, A.; Brolo, A. G.; Choo, J.; Cialla-May, D.; Deckert, V.; Fabris, L.; Faulds, K.; de Abajo, F. J. G., Goodacre, R.; Graham, D.; Haes, A. J.; Haynes, C. L.; Huck, C.; Itoh, T.; Käll, M.; Kneipp, J.; Kotov, N. A.; Kuang, H.; Le Ru, E. C.; Lee, H. K.; Li, J.; Ling, X. Y.; Maier, S. A.; Mayerhöfer, T.; Moskovits, M.; Murakoshi, K.; Nam, J.; Nie, S.; Ozaki, Y.; Pastoriza-Santos, I.; Perez-Juste, J.; Popp, J.; Pucci, A.; Reich, S.; Ren, B.; Schatz, G. C.; Shegai, T.; Schlücker, S.; Tay, L.; Thomas, K. G.; Tian, Z.; Van Duyne, R. P.; Vo-Dinh, T.; Wang, Y.; Willets, K. A.; Xu, C.; Xu, H.; Xu, Y.; Yamamoto, Y. S.; Zhao, B.; Liz-Marzán, L. M. Present and future of surface-enhanced Raman scattering. *ACS Nano* **2020**, *14*, 28–117.

[7] Alessandri, I.; Lombardi, J. R.; Enhanced Raman scattering with dielectrics. *Chem. Rev.* **2016**, *116*, 14921–14981.

[8] Lan, L. L.; Gao, Y. M.; Fan, X. C.; Li, M. Z.; Hao, Q.; Qiu, T. The origin of ultrasensitive SERS sensing beyond plasmonics. *Front. Phys.* **2021**, *16*, 43300.

[9] Lombardi, J. R.; Birke, R. L. Theory of surface-enhanced Raman scattering in semiconductors. *J. Phys. Chem. C* **2014**, *118*, 11120–11130.

[10] Xu, J. T.; Li, X. T.; Wang, Y. X.; Guo. R. H.; Shang S. M.; Jiang, S. X. Flexible and reusable cap-like thin $Fe_2O_3$ film for SERS applications. *Nano Res.* **2019**, *12*, 381–388.

[11] Song, G.; Gong, W. B.; Cong, S.; Zhao, Z. G. Ultrathin two-dimensional nanostructures: surface defects for morphology-driven enhanced semiconductor SERS. *Angew. Chem. Int. Ed.* **2021**, *60*, 5505–5511.

[12] Wang, X. T.; Shi, W. X.; Wang, S. X.; Zhao, H. W.; Lin, J.; Yang, Z.; Chen, M.; Guo, L. Two-dimensional amorphous $TiO_2$ nanosheets enabling high-efficiency photoinduced charge transfer for excellent SERS activity. *J. Am. Chem. Soc.* **2019**, *141*, 5856-5862.

[13] Fan, X. C.; Li, M. Z.; Hao, Q.; Zhu, M. S.; Hou, X. Y.; Huang, H.; Ma, L. B.; Schmidt, O. G.; Qiu, T. High SERS sensitivity enabled by synergistically enhanced photoinduced charge transfer in amorphous nonstoichiometric semiconducting films. *Adv. Mater. Interfaces* **2019**, *6*, 1901133.

[14] Wang, X. T.; Shi, W. X.; Jin, Z.; Huang, W. F.; Lin, J.; Ma, G. S.; Li, S. Z.; Guo, L. Remarkable SERS activity observed from amorphous ZnO nanocages. *Angew. Chem., Int. Ed.* **2017**, *56*, 9851–9855.

[15] Zheng, X. L.; Guo, H. L.; Xu, Y.; Zhang, J. L.; Wang, L. Z. Improving SERS sensitivity of $TiO_2$ by utilizing the heterogeneity of facet-potentials. *J. Mater. Chem. C* **2020**, *8*, 13836–13842.

[16] Lin, J.; Hao, W.; Shang, Y.; Wang, X. T.; Qiu, D. L.; Ma, G. S.; Chen, C.; Li, S. Z.; Guo, L. Direct experimental observation of facet-dependent SERS of $Cu_2O$ polyhedra. *Small* **2018**, *14*, 1703274.

[17] Lin, J., Shang, Y.; Li, X. X.; Yu, J.; Wang, X. T.; Guo, L. Ultrasensitive SERS detection by defect engineering on Single $Cu_2O$ superstructure particle. *Adv. Mater.* **2017**, *29*, 1604797.

[18] Yang, L. L.; Peng, Y. S.; Yang, Y.; Liu, J. J.; Huang, H. L.; Yu. B H., Zhao, J. M.; Lu, Y. L.; Huang, Z. R.; Li, Z. Y.; Lombardi, J. R. A novel ultra-sensitive semiconductor SERS substrate boosted by the coupled resonance effect. *Adv. Sci.* **2019**, *6*, 1900310.

[19] Lee, Y.; Kim, H.; Lee, J.; Yu, S. H.; Hwang, E.; Lee C.; Ahn, J. H.; Cho, J. H.; Enhanced Raman scattering of rhodamine 6G films on two-dimensional transition metal dichalcogenides correlated to photoinduced charge transfer. *Chem. Mater.* **2016**, *28*, 180–187.

[20] Miao, P.; Qin, J. K.; Shen, Y. F.; Su, H. M.; Dai, J. F.; Song, B.; Du, Y. C.; Sun, M. T.; Zhang, W.; Wang, H. L.; Xu, C. Y.; Xu, P.; Unraveling the Raman enhancement mechanism on1T′-phase $ReS_2$ nanosheets. *Small* **2018**, *14*, 1704079.

[21] Li, M. Z.; Gao, Y. M.; Fan, X. C.; Wei, Y. J.; Qiu, T. Origin of layer-dependent SERS tunability in 2D transition metal dichalcogenides. *Nanoscale Horiz.* **2021**, *6*, 186–191.

[22] Lee, Y.; Kim, H.; Lee, J. B.; Cho, J. H.; Ahn, J. H. Pressure-induced chemical enhancement in Raman scattering from graphene-Rhodamine 6G-graphene sandwich structures. *Carbon* **2015**, *89*, 318–327.

[23] Sun, H. H.; Yao, M. G.; Liu, S.; Song, Y. P.; Shen, F. R.; Dong, J. J.; Yao, Z.; Zhao, B.; Liu, B. B. SERS selective enhancement on monolayer $MoS_2$ enabled by a pressure-induced shift from resonance to charge transfer. *ACS Appl. Mater.*

*Interfaces* **2021**, *13*, 26551–26560.

[24] Hou, X. Y.; Zhang, X. Y.; Ma, Q. W.; Tang, X.; Hao, Q.; Cheng, Y. C.; Qiu, T. Alloy engineering in few-layer manganese phosphorus trichalcogenides for surface-enhanced Raman scattering. *Adv. Funct. Mater.* **2020**, *30*, 1910171.

[25] Zheng, Z. H.; Cong, S.; Gong, W. B.; Xuan, J. N.; Li, G. H.; Lu, W. B.; Geng, F. X.; Zhao, Z. G. Semiconductor SERS enhancement enabled by oxygen incorporation. *Nat. Commun.* **2017**, *8*, 1993.

[26] Cong, S.; Yuan, Y. Y.; Chen, Z. G.; Hou, J. Y.; Yang, M.; Su, Y. L.; Zhang, Y. Y.; Li, L.; Li, Q. W.; Geng, F. X.; Zhao, Z. G. Noble metal-comparable SERS enhancement from semiconducting metal oxides by making oxygen vacancies. *Nat. Commun.* **2015**, *6*, 7800.

[27] Lan, L. L.; Hou, X. Y.; Gao, Y. M.; Fan, X. C.; Qiu, T. Inkjet-printed paper-based semiconducting substrates for surface-enhanced Raman spectroscopy. *Nanotechnology* **2020**, *31*, 055502.

[28] Ling, X.; Fang, W. J.; Lee, Y. H.; Araujo, P. T.; Zhang, X.; Rodriguez-Nieva, J. F.; Lin, Y. X.; Zhang, J.; Kong, J.; Dresselhaus, M. S. Raman Enhancement Effect on Two-Dimensional Layered Materials: Graphene, h-BN and MoS$_2$. *Nano Lett.* **2014**, *14*, 3033–3040.

[29] Kannan, P. K.; Shankar, P.; Blackman, C.; Chung, C. H. Recent advances in 2D inorganic nanomaterials for SERS Sensing. *Adv. Mater.* **2019**, *31*, 1803432.

[30] Demirel, G.; Gieseking, R. L. M.; Ozdemir, R.; Kahmann, S.; Loi, M. A.; Schatz, G. C.; Facchetti, A.; Usta, H. Molecular engineering of organic semiconductors enables noble metal-comparable SERS enhancement and sensitivity. *Nat. Commun.* **2019**, *10*, 5502.

[31] Yilmaz, M.; Babur, E.; Ozdemir, M.; Gieseking, R. L.; Dede, Y.; Tamer, U.; Schatz, G. C.; Facchetti, A.; Usta, H.; Demirel, G. Nanostructured organic semiconductor films for molecular detection with surface-enhanced Raman spectroscopy. *Nat. Mater.* **2017**, *16*, 918–925.

[32] Sun, H. Z.; Cong, S.; Zheng, Z. H.; Wang, Z.; Chen, Z. G.; Zhao, Z. G. Metal-organic frameworks as surface enhanced Raman scattering substrates with high tailorability. *J. Am. Chem. Soc.* **2019**, *141*, 870–878.

[33] Su, X. Y.; Ma, H.; Wang, H.; Li, X. L.; Han, X. X.; Zhao, B. Surface-enhanced Raman scattering on organic-inorganic hybrid perovskites. *Chem. Commun.* **2018**, *54*, 2134–2137.

[34] Fan, X. C.; Hao, Q.; Qiu, T.; Chu, P. K. Improving the performance of light-emitting diodes via plasmonic-based strategies. *J. Appl. Phys.* **2020**, *127*, 040901.

[35] Agrawal, A.; Cho, S. H.; Zandi, O.; Ghosh, S.; Johns, R. W.; Milliron, D. J. Localized surface plasmon resonance in semiconductor nanocrystals. *Chem. Rev.* **2018**, *118*, 3121–3207.

[36] Li, P.; Zhu, L.; Ma, C.; Zhang, L. X.; Guo, L.; Liu, Y. W.; Ma, H.; Zhao, B. Plasmonic molybdenum tungsten oxide hybrid

with surface-enhanced Raman scattering comparable to that of noble metals. *ACS Appl. Mater. Interfaces* **2020**, *12*, 19153–19160.

[37] Hou, X. Y.; Luo, X. G.; Fan, X. C.; Peng, Z. H.; Qiu, T. Plasmon-coupled charge transfer in $WO_{3-x}$ semiconductor nanoarrays: toward highly uniform silver-comparable SERS platforms. *Phys. Chem. Chem. Phys.* **2019**, *21*, 2611–2618.

[38] Liu, W.; Bai, H.; Li, X. S.; Li, W. T.; Zhai, J. F.; Li, J. F.; Xi, G. C. Improved surface-enhanced Raman spectroscopy sensitivity on metallic tungsten oxide by the synergistic effect of surface plasmon resonance coupling and charge transfer. *J. Phys. Chem. Lett.* **2018**, *9*, 4090–4100.

[39] Tan, X. J.; Wang, L. Z.; Cheng, C.; Yan, X. F.; Shen, B.; Zhang, J. L. Plasmonic $MoO_{3-x}@MoO_3$ nanosheets for highly sensitive SERS detection through nanoshell-isolated electromagnetic enhancement. *Chem. Commun.* **2016**, *52*, 2893–2896.

[40] Lan, L. L.; Fan, X. C.; Gao, Y. M.; Li, G. Q.; Hao, Q.; Qiu, T. Plasmonic metal carbide SERS. *J. Mater. Chem. C* **2020**, *8*, 14523–14530.

[41] Ye, Y. T.; Yi, W. C.; Liu, W.; Zhou, Y.; Bai, H.; Li, J. F.; Xi, G. C. Remarkable surface-enhanced Raman scattering of highly crystalline monolayer $Ti_3C_2$ nanosheets. *Sci. China Mater.* **2020**, *63*, 794–805.

[42] Zhong, Y.; Xia, X. H.; Shi, F.; Zhan, J. Y.; Tu, J. P.; Fan, H. J. Transition metal carbides and nitrides in energy storage and conversion. *Adv. Sci.* **2016**, *3*, 1500286.

[43] Huang, W. C.; Gao, Y.; Wang, J. X.; Ding, P. C.; Yan, M.; Wu, F. M.; Liu, J.; Liu, D. Q.; Guo, C. S.; Yang, B.; Cao, W. W. Plasmonic enhanced reactive oxygen species activation on low-work-function tungsten nitride for direct near-infrared driven photocatalysis. *Small* **2020**, *16*, 2004557.

[44] Yu, L.; Zhu, Q.; Song, S. W.; McElhenny, B.; Wang, D. Z.; Wu, C. Z.; Qin, Z. J.; Bao, J. M.; Yu, Y.; Chen, S.; Ren, Z. F. Non-noble metal-nitride based electrocatalysts for high-performance alkaline seawater electrolysis. *Nat. Commun.* **2019**, *10*, 5106.

[45] Hao, Q.; Li, W.; Xu, H. Y.; Wang, J. W.; Yin, Y.; Wang, H. Y.; Ma, L. B.; Ma, F.; Jiang, X. C.; Schmidt, O. G.; Chu, P. K. $VO_2$/TiN plasmonic thermochromic smart coatings for room-temperature applications. *Adv. Mater.* **2018**, *30*, 1705421.

[46] Guan, H. M.; Yi, W. C.; Li, T.; Li, Y. H.; Li, J. F.; Bai, H.; Xi G. C. Low temperature synthesis of plasmonic molybdenum nitride nanosheets for surface enhanced Raman scattering. *Nat. Commun.* **2020**, *11*, 3889.

[47] Soundiraraju, B.; George, B. K. Two-dimensional titanium nitride ($Ti_2N$) MXene: synthesis, characterization, and potential application as surface-enhanced Raman scattering substrate. *ACS Nano* **2017**, *11*, 8892–8900.

[48] Guan, H. M.; Li, W. T.; Han, J.; Yi, W. C.; Bai, H.; Kong, Q. H.; Xi, G. C. General molten-salt route to three-dimensional porous transition metal nitrides as sensitive and stable Raman substrates. *Nat. Commun.* **2021**, *12*, 1376.

[49] Esmaeilzadeh, M.; Dizajghorbani-Aghdam, H.; Malekfar, R. Surface-enhanced Raman scattering of methylene blue on

titanium nitride nanoparticles synthesized by laser ablation in organic solvents. *Spectrochim. Acta A* **2021**, *257*, 119721.

[50] Du, R. F.; Yi, W. C.; Li, W. T.; Yang, H. F.; Bai, H.; Li, J. F.; Xi, G. C. Quasi-metal microwave route to MoN and $Mo_2C$ ultrafine nanocrystalline hollow spheres as surface-enhanced Raman scattering substrates. *ACS Nano* **2020**, *14*, 13718–13726.

[51] Zhu, Y. P.; Chen, G.; Zhong, Y. J.; Zhou, W.; Shao, Z. P. Rationally designed hierarchically structured tungsten nitride and nitrogen-rich graphene-like carbon nanocomposite as efficient hydrogen evolution electrocatalyst. *Adv. Sci.* **2018**, *5*, 1700603.

[52] Karaballi, R. A.; Humagain, G.; Fleischman, B. R. A.; Dasog, M. Synthesis of plasmonic group-4 nitride nanocrystals by solid-state metathesis. *Angew. Chem. Int. Edit.* **2019**, *58*, 3147–3150.

[53] Yang, X. B.; Aydin, E.; Xu, H.; Kang, J. X.; Hedhili, M.; Liu, W. Z.; Wan, Y. M.; Peng, J.; Samundsett, C.; Cuevas, A.; Wolf, S. D. Tantalum nitride electron-selective contact for crystalline silicon solar cells. *Adv. Energy Mater.* **2018**, *8*, 1800608.

[54] Kim, G. T.; Park, T. K.; Chung, H.; Kim, Y. T.; Kwon, M. H.; Choi, J. G. Growth and characterization of chloronitroaniline crystals for optical parametric oscillators I. XPS study of Mo-based compounds. *Appl. Surf. Sci.* **1999**, *152*, 35–43.

[55] Joy, V. T.; Srinivasan, T. K. K. Fourier-transform surface enhanced Raman scattering study on thiourea and some substituted thioureas adsorbed on chemically deposited silver films. *Spectrochim. Acta A* **1999**, *55*, 2899–2909.

[56] Song, X. J.; Wang, Y.; Zhao, F.; Li, Q. C.; Ta, H. Q.; Ruemmeli, M. H.; Tully, C. G.; Li, Z. Z.; Yin, W. J.; Yang, L. T.; Lee, K. B.; Yang, J.; Bozkurt, I.; Liu, S. W.; Zhang, W. J.; Chhowalla, M. Plasmon-free surface-enhanced Raman spectroscopy using metallic 2D materials. *ACS Nano* **2019**, *13*, 8312–8319.

[57] Tao, L.; Chen, K.; Chen, Z. F.; Cong, C. X.; Qiu, C. Y.; Chen, J. J.; Wang, X. M.; Chen, H. J.; Yu, T.; Xie, W. G.; Deng, S. Z.; Xu, J. B. 1T′ transition metal telluride atomic layers for plasmon-free SERS at femtomolar levels. *J. Am. Chem. Soc.* **2018**, *140*, 8696–8704.

[58] Hu, Y. M.; Li, J. Y.; Chen, N. Y.; Chen, C. Y.; Han, T. C.; Yu, C. C. Effect of sputtering power on crystallinity, intrinsic defects, and optical and electrical properties of Al-doped ZnO transparent conducting thin films for optoelectronic devices. *J. Appl. Phys.* **2017**, *121*, 085302.

[59] Tsetseris, L.; Kalfagiannis, N.; Logothetidis, S.; Pantelides, S. T. Structure and interaction of point defects in transition-metal nitrides. *Phys. Rev. B* **2007** *76*, 224107.

[60] Tsetseris, L.; Kalfagiannis, N.; Logothetidis, S.; Pantelides, S. T. Trapping and release of impurities in TiN: A first-principles study. *Phys. Rev. B* **2008** *78*, 094111.

[61] Pei, C. R.; Deng, L. J.; Xiang, C. J.; Zhang, X.; Sun, D. Effect of the varied nitrogen vacancy concentration on mechanical and electrical properties of $ZrN_x$ thin films. *Thin Solid Films* **2019**, *683*, 57–66.

[62] Yu, J.; Yang, M. S.; Li, Z.; Liu, C. D.; Wei, Y. S.; Zhang, C., Man, B. Y.; Lei, F. C. Hierarchical particle-in-quasicavity architecture for ultratrace in situ Raman sensing and its application in real-time monitoring of toxic pollutants. *Anal. Chem.* **2020**, *92*, 14754–14761.

[63] Li, H.; Zhang, Q.; Yap, C. C. R.; Tay, B. K.; Edwin, T. H. T.; Olivier, A.; Baillargeat, D. From bulk to monolayer $MoS_2$: evolution of Raman scattering. *Adv. Funct. Mater.* **2012**, *22*, 1385–1390.

[64] Canamares, M. V.; Chenal, C.; Birke, R. L.; Lombardi, J. R. DFT, SERS and single-molecule SERS of crystal violet. *J. Phys. Chem. C* **2008**, *112*, 2028–20300.

[65] Zong, C.; Chen, C. J.; Wang, X.; Hu, P.; Liu, G. K.; Ren, B. Single-Molecule Level Rare Events Revealed by Dynamic Surface-Enhanced Raman Spectroscopy. *Anal. Chem.* **2020**, *92*, 15806–15810.

[66] Chen, C. J.; Zong, C.; Liu, G. K.; Ren, B. Adsorption behavior of rhodamine 6G on silver surfaces studied by electrochemical surface-enhanced Raman spectroscopy. *J. Electrochem.* **2016**, *22*, 32–26.

[67] Muehlethaler, C.; Considine, C. R.; Menon, V; Lin, W. C.; Lee, Y. H.; Lombardi, J. R. Ultrahigh Raman enhancement on monolayer $MoS_2$. *ACS Photonics* **2016**, *3*, 1164–1169.

[68] Zheng, W. H.; Zheng, B. Y.; Yan, C. L.; Liu, Y.; Sun, X. X.; Qi, Z. Y.; Yang, T. F.; Jiang, Y.; Huang, W.; Fan, P.; Jiang, F.; Ji, W.; Wang, X.; Pan, A. L. Direct vapor growth of 2D vertical heterostructures with tunable band alignments and interfacial charge transfer behaviors. *Adv. Sci.* **2019**, *6*, 1802204.

[69] Li, M. Z.; Fan, X. C.; Gao, Y. M.; Qiu, T. $W_{18}O_{49}$/Monolayer $MoS_2$ heterojunction-enhanced Raman scattering. *J. Phys. Chem. Lett.* **2019**, *10*, 4038–4044.

[70] Yin, Y.; Miao, P.; Zhang, Y. M.; Han, J. C.; Zhang, X. H.; Gong, Y.; Gu, L.; Xu, C. Y.; Yao, T.; Xu, P.; Wang, Y.; Song, B.; Jin, S. Significantly increased Raman enhancement on $MoX_2$ (X = S, Se) monolayers upon phase Transition. *Adv. Funct. Mater.* **2017**, *27*, 1606694.

# Graphical Table of Contents

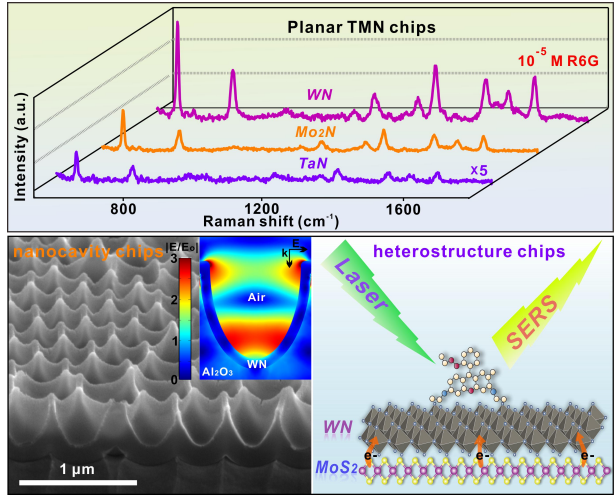

**Text:** We propose a general strategy to fabricate a series of planar transition-metal nitride SERS chips. Further, structural engineering of these TMN chips is used to improve their SERS activity by constructing nanocavity and heterostructure, the Raman EFs could be greatly improved to ∼$10^7$.

## Section S1: Fabrication of the Ag chips

Ag chips were prepared on Si wafer in an argon (40 sccm) atmosphere using a direction current magnetron system at room temperature [1]. A high purity Ag plate (99.99%, 50mm-diameter) was used as a target. The power was 80 W and the sputtered time was 240 s with a pressure of 1.75 Pa.

## Section S2: Calculation of the enhancement factor (EF)

The EF of the samples could be estimated using the equation below [2]:

$$EF = (I_{SERS}/N_{SERS}) / (I_{bulk}/N_{bulk}) \qquad (1)$$

$$N_{SERS} = CVN_A A_{Raman}/A_{Sub} \qquad (2)$$

$$N_{bulk} = \rho h A_{Raman} N_A / M \qquad (3)$$

where $I_{SERS}$ and $I_{bulk}$ are the intensities of the selected Raman peak in the SERS and non-SERS spectra, respectively. $N_{SERS}$ and $N_{bulk}$ are the average number of molecules in scattering area for SERS and non-SERS measurement. The data of bulk R6G is used as non-SERS-active reference. $C$ is the molar concentration of R6G solution and $V$ is the volume of the droplet (20 μL). $N_A$ is Avogadro constant (6.023 × $10^{23}$ mol$^{-1}$). $A_{Raman}$ is the laser spot area. $A_{Sub}$ is the effective area of the substrate, which is approximately 9 πmm$^2$. The confocal depth $h$ of the laser beam is 21 μm. The molecular weight $M$ of R6G is 479 g mol$^{-1}$ and density $\rho$ of bulk R6G is 1.15 g cm$^{-3}$.

## Section S3: Calculation of the degree of charge transfer

The degree of charge transfer ($P_{CT}(k)$) is used for investigating the charge transfer (CT) contributions to the SERS intensity, which can be described as follow formula [3-4]:

$P_{CT}(k) = (I^k(CT)-I^k(SPR)) / (I_k(CT)-I_0(SPR))$

where $I^k(CT)$ is the Raman intensity of a band in which the CT contributes to the intensity of enhanced Raman signals. Two bands with no CT contribution are selected as a reference. One is $I^0(SPR)$, the intensity of the totally symmetric band only comes from the contribution of the surface plasmon resonance (SPR). The other is called $I^k(SPR)$. If the k band is totally symmetric vibration band, the main contribution to the intensity arises from the SPR, and $I^k(SPR) = I^0(SPR)$. If k band is non-totally symmetric vibration band, the contribution from CT dominates the intensity. At this moment, $I^k(SPR)$ is usually quite small, and it can be assumed to be zero in many cases. It can be seen from formula that when $P_{CT}$ is zero, there are no CT contributions; when $P_{CT}$ close to 1, the CT contributions will dominate the spectrum. We could select the totally symmetric 1360 cm$^{-1}$ mode and the non-totally symmetric 612 cm$^{-1}$ mode as line 0 and k, respectively [5]. Ignoring the adsorption behavior of molecules, the degrees of CT for R6G (1×10$^{-5}$ M) on WN, TaN and Mo$_2$N are 0.65, 0.71 and 0.66 respectively, which demonstrates that the chemical mechanism is the main enhancement mechanism.

**Section S4: First-principles calculations**

All our first-principles calculations were carried out within the framework of density functional theory (DFT) as implemented in the Vienna ab initio simulation package (VASP), to obtain insights into the equilibrium structure and electronic properties of WN, TaN, and Mo2N. Periodic boundary conditions are used for the 3D structures in the process of optimizing the equilibrium structure and calculating the band structures. The reciprocal space is sampled by a fine grid of 8 × 8 × 8 k points in the Brillouin zone. The average electrostatic potential study of WN, TaN, and Mo$_2$N adopted the 2D periodic boundary structures with 2 × 2 supercell and a vacuum spacing of 30 Å was employed to eliminate the interactions between layers. A k-point sampling set of 8×8×1 was tested to be converged. We used projector augmented-wave (PAW) pseudopotentials and the Perdew-Burke-Ernzerhof (PBE) exchange-correlation approximation. The electronic wave function is expanded on a plane-wave basis set with a cutoff energy of 450 eV. A total energy difference between subsequent self-consistency iterations below 10$^{-5}$ eV is used as the criterion for reaching self-consistency. All geometries have been optimized using the conjugate-gradient method, until none of the residual Hellmann-Feynman forces exceeded 10$^{-2}$ eV/Å.

**Section S5: Simulation of the electromagnetic field distribution**

The finite element method (FEM) simulations are carried out on COMSOL. These modes are excited by a plane wave source with incident direction perpendicular to the plane of structure. The incident light wavelength is 532 nm, and its electric field intensity is 1 V/m. The scattering boundary conditions are applied to the model. The dielectric constants of tungsten nitride are set as $\varepsilon' = -8.80$ and $\varepsilon'' = 1.15$ [6]. The diameter (D), thickness (T), and depth (H) of the nanocavities were set to 370 nm, 40 nm, and 360 nm, respectively. The planar structure was modeled as a tungsten nitride layer (40 nm) coated on Si wafer.

**Section S6: Deposition rate of the WN films**

Constant technological parameters (working pressure, gas flow, sputtering power and reaction temperature) during the magnetron sputtering determine the growth velocity of the chips. We set the sputtering time as 7000 s and obtained ~1.175-μm-thick film confirmed by SEM (Figure S14), which shows that the deposition rate is about 0.168 nm s$^{-1}$ (1.175 μm ÷ 7000 s = 0.168 nm s$^{-1}$) under the circumstances.

**Section S7: Figures**

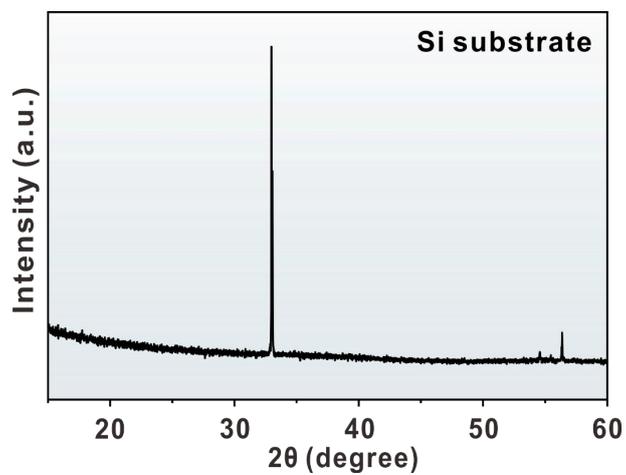

**Figure S1** The XRD pattern of Si substrate.

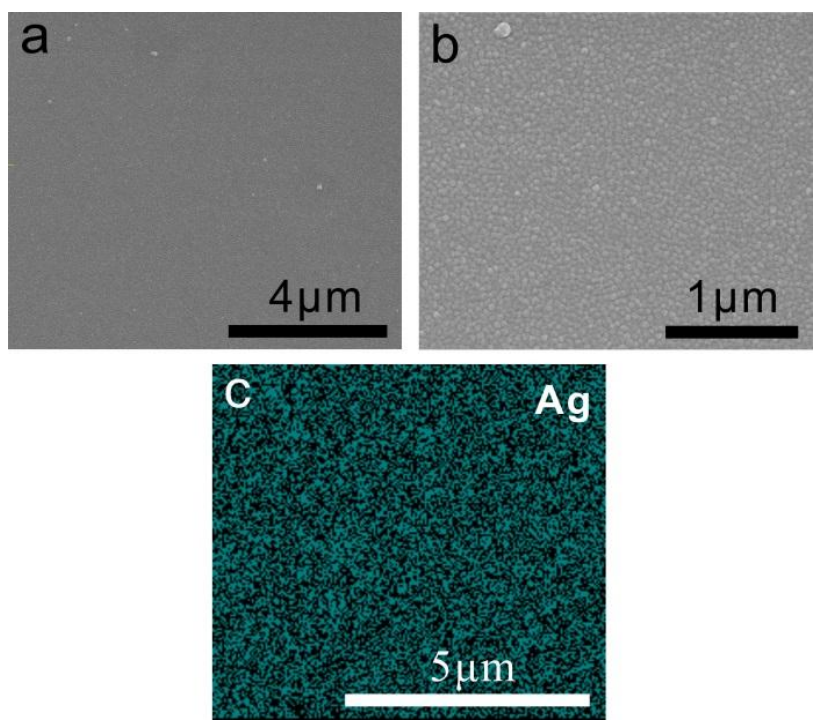

**Figure S2** (a) Low-magnification SEM image, (b) high-magnification SEM image, and (c) elemental mapping image of Ag chips.

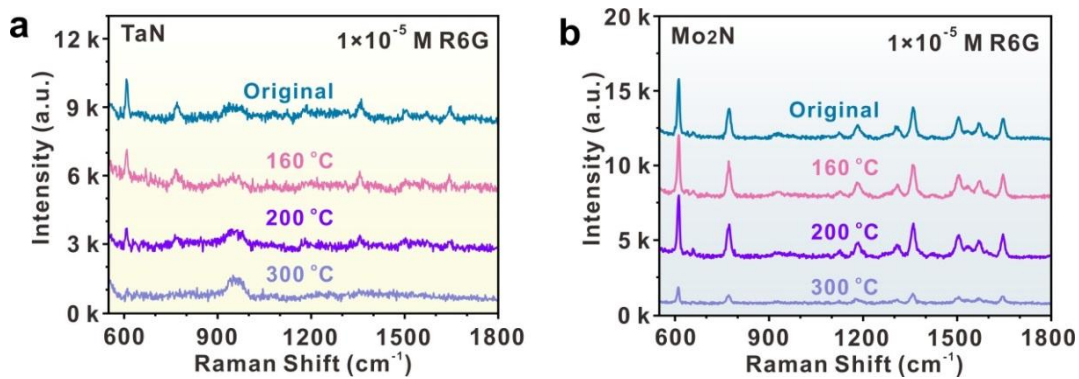

**Figure S3** (a) The Raman spectra of R6G ($10^{-5}$ M) adsorbed on original and high temperature treated TaN chips. (b) The Raman spectra of R6G ($10^{-5}$ M) adsorbed on original and high temperature treated $Mo_2N$ chips.

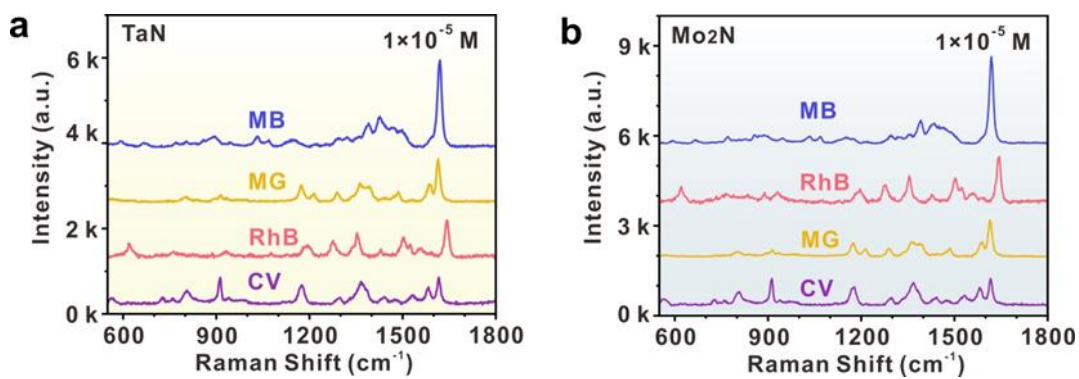

**Figure S4** The Raman spectra of a series of organic compounds ($10^{-5}$ M) adsorbed on (a) TaN and (b) $Mo_2N$ chips.

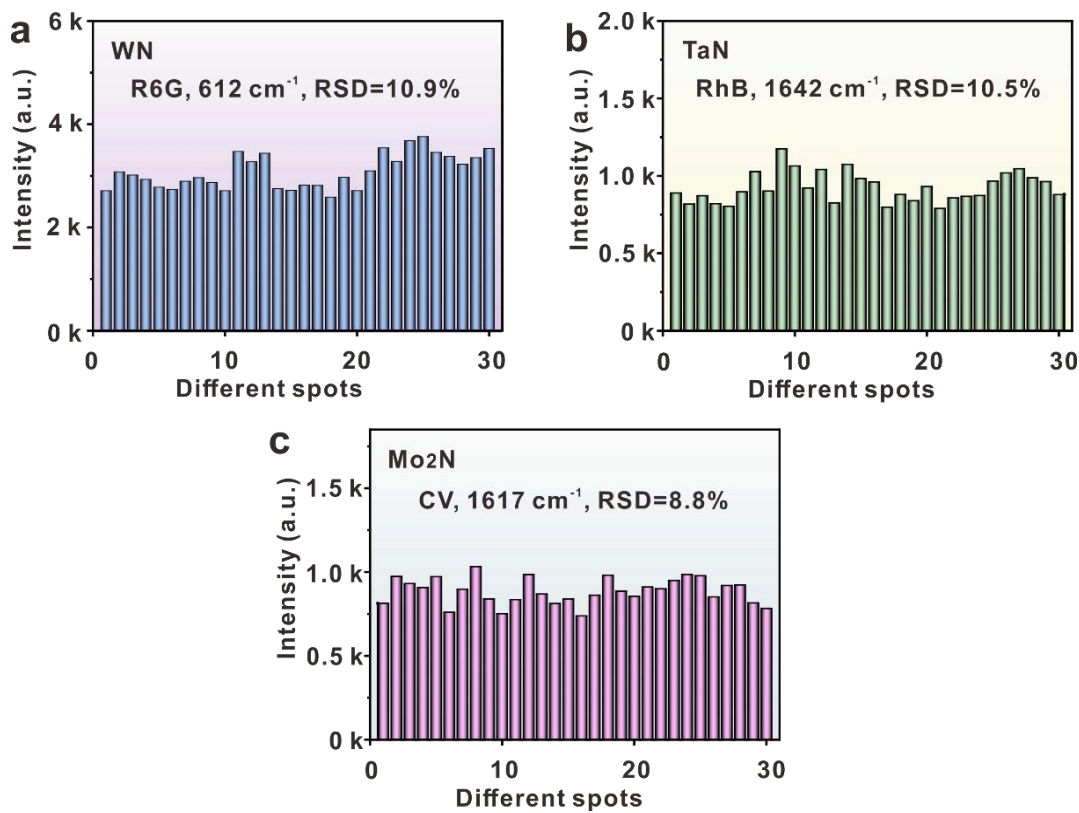

**Figure S5** The RSD values calculated for the characteristic peaks at (a) 612 cm$^{-1}$ (10$^{-6}$ M R6G adsorbed on WN), (b) 1642 cm$^{-1}$ (10$^{-5}$ M RhB adsorbed on TaN), and (c) 1642 cm$^{-1}$ (10$^{-5}$ M CV adsorbed on Mo$_2$N), respectively.

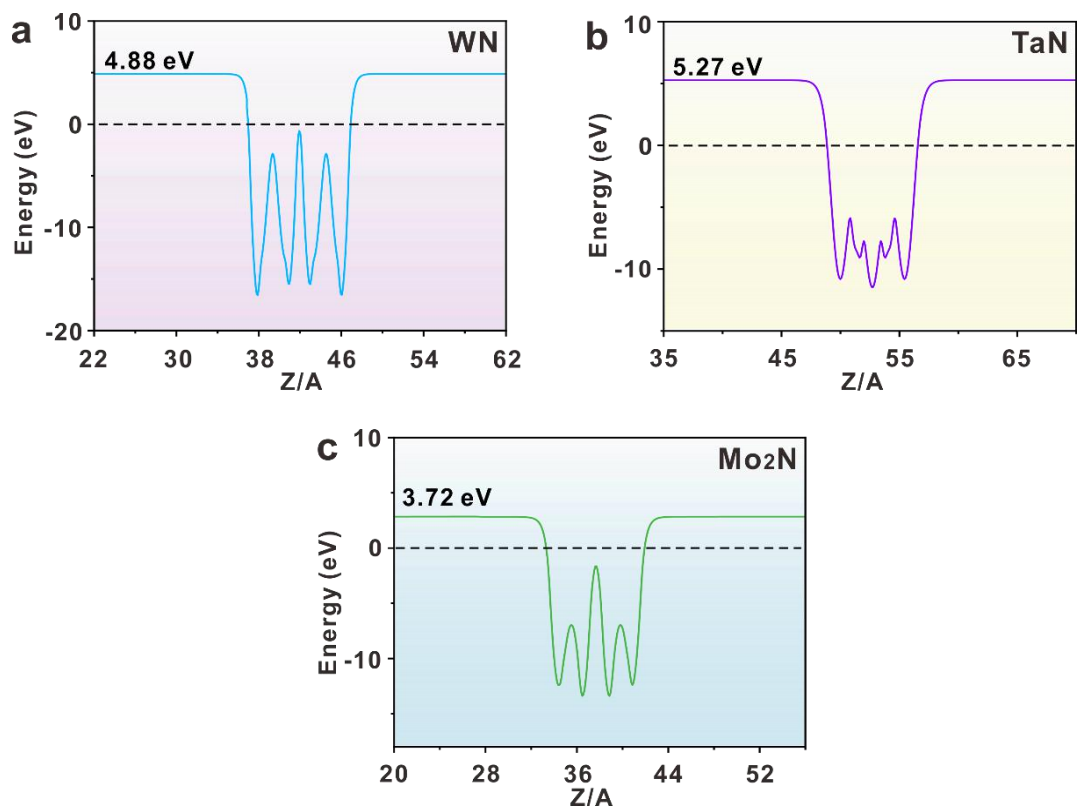

**Figure S6** The DFT calculated work function of cubic (a) WN, (b) TaN, and (c) Mo$_2$N.

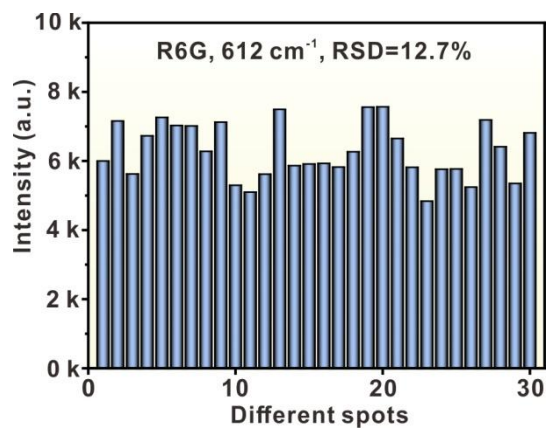

**Figure S7** The RSD value (612 cm$^{-1}$) of R6G (10$^{-6}$ M) gathered from 30 random sites on the WN nanocavity.

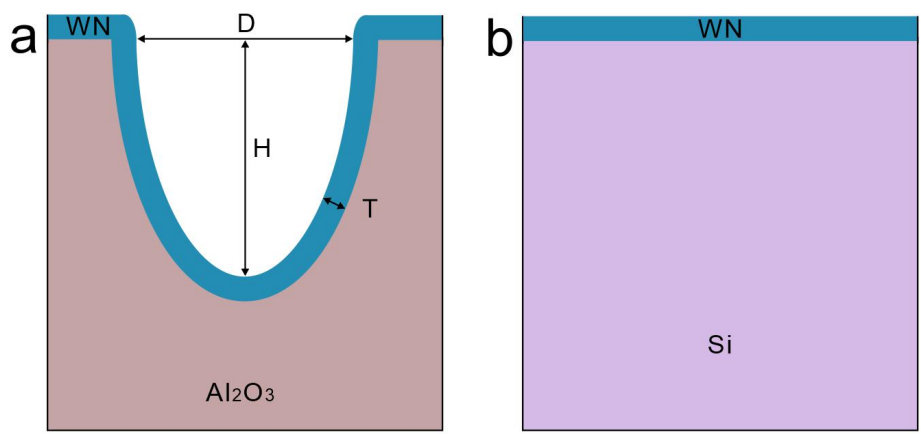

**Figure S8** The structural models of (a) nanocavity and (b) planar chips utilized in the FEM simulations.

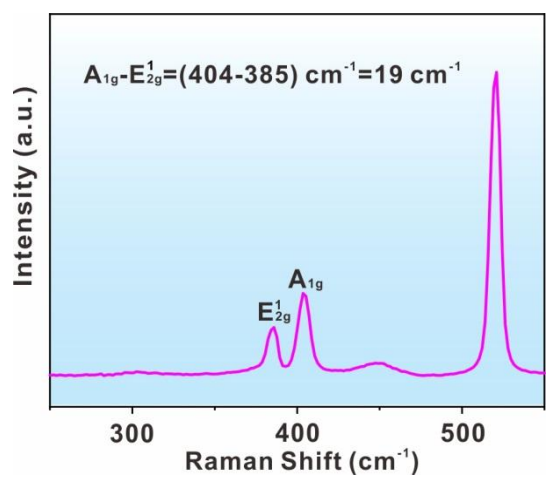

**Figure S9** The Raman spectrum of monolayer MoS$_2$.

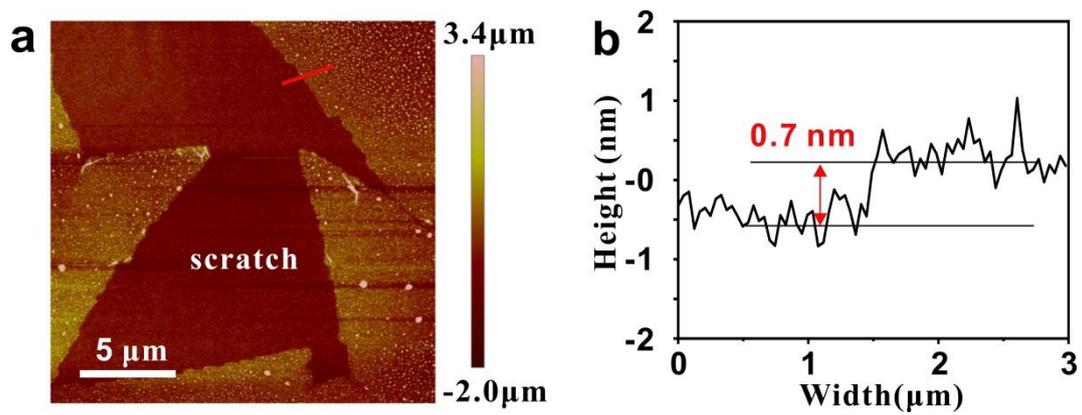

**Figure S10** (a) The AFM image of monolayer MoS$_2$, and (b) section analysis along the red line in (a).

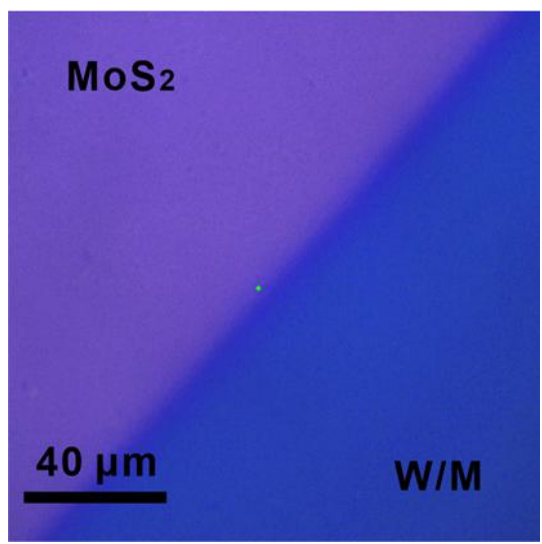

**Figure S11** The optical photograph of WN/monolayer MoS$_2$ chips.

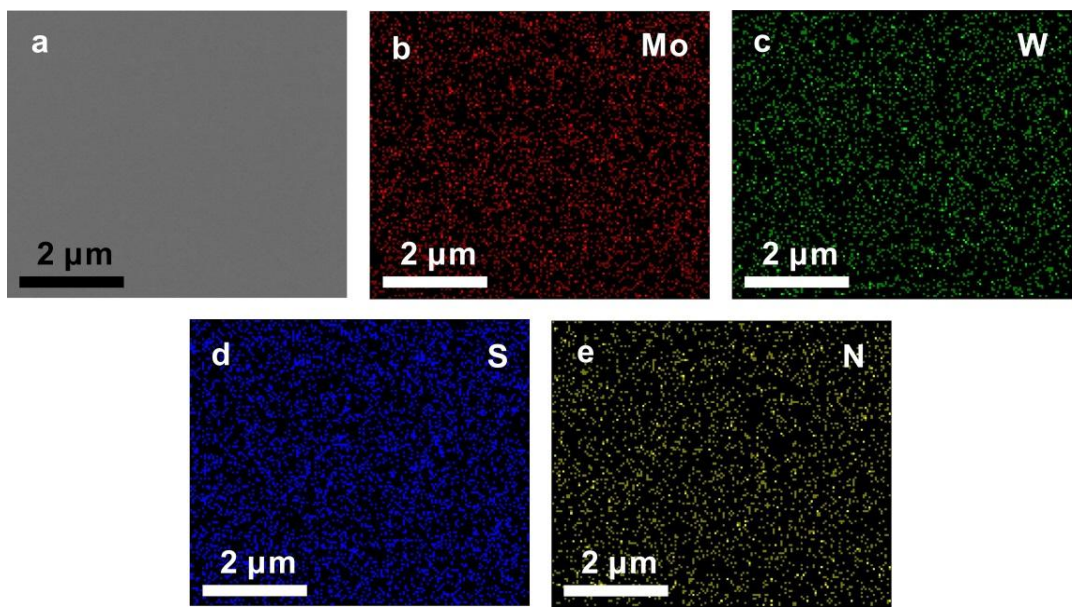

**Figure S12** (a) The SEM image and (b-e) corresponding elemental mapping images of WN/monolayer MoS$_2$ chips.

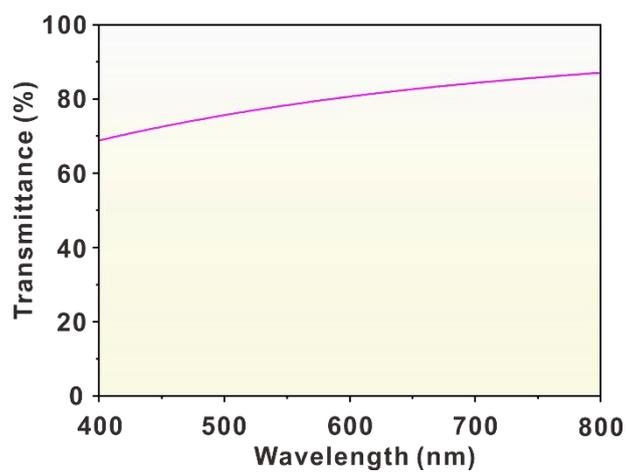

**Figure S 13** The transmittance of (10 nm) WN layer deposited on transparent substrate.

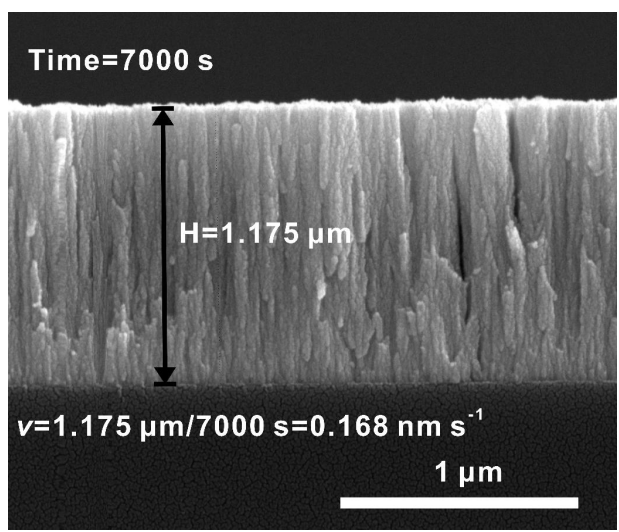

**Figure S14** Cross-sectional SEM image for WN chips prepared by sputtering (sputtering time is 7000 s).

**Section S8: Tables**

| Sample | Target (purity) | Pressure (Pa) | Temperature (°C) | Power (W) | Ar: N2 | Time (s) |
|---|---|---|---|---|---|---|
| TaN | Ta (99.99%) | 1.0 | 25 | 80 | 27:3 | 240 |
| Mo$_2$N | Mo (99.99%) | 0.7 | 25 | 80 | 20:6 | 240 |

**Table S1** Sputtering parameters of TaN and Mo$_2$N chips.

**References**

[1] Qiu, T.; Zhang, W. J.; Lang, X. Z.; Zhou, Y. J.; Cui, T. J.; Chu, P. K. Controlled assembly of highly Raman-enhancing silver nanocap arrays templated by porous anodic alumina membranes. *Small* **2009**, *5*, 2333–2337.

[2] Cong, S.; Yuan, Y. Y.; Chen, Z. G.; Hou, J. Y.; Yang, M.; Su, Y. L.; Zhang, Y. Y.; Li, L.; Li, Q. W.; Geng, F. X.; Zhao, Z. G. Noble metal-comparable SERS enhancement from semiconducting metal oxides by making oxygen vacancies. *Nat. Commun*. **2015**, *6*, 7800.

[3] Canamares, M. V.; Chenal, C.; Birke, R. L.; Lombardi, J. R. DFT, SERS, and single-molecule SERS of crystal violet. *J. Phys. Chem. C* **2008**, 112, 20295–20300.

[4] Richter, A. P.; Lombardi, J. R.; Zhao, B. Size and Wavelength Dependence of the Charge-Transfer Contributions to Surface-Enhanced
 Raman Spectroscopy in Ag/PATP/ZnO Junctions. *J. Phys. Chem. C* **2010**, *114*, 1610–1614.

[5] Li, M.; Wei, Y.; Fan, X.; Li, G., Hao, Q.; Qiu, T. Mixed-dimensional van der Waals heterojunction-enhanced Raman scattering. *Nano Res.* 2021, doi.org/10.1007/s12274-021-3537-2.

[6] Kumar, M.; Umezawa, N.; Ishii, S.; Nagao, T. Examining the performance of refractory conductive ceramics as plasmonic materials: a theoretical approach. *ACS photonics* **2016**, *3*, 43–50.